\documentclass[12pt]{iopart}
\usepackage{graphicx}
\usepackage{setspace}
\usepackage{amsfonts}
\usepackage{amssymb,color}

\usepackage{graphicx,color}
\usepackage{ifpdf}
\usepackage[latin1]{inputenc}

\begin{document}

\title{Effects of diffusion in competitive contact processes on 
bipartite lattices}

\author{M. M. de Oliveira$^{1}$ and C. E. Fiore$^2$}
\address{
$^1$Departamento de F\'{\i}sica e Matem\'atica,
CAP, Universidade Federal de S\~ao Jo\~ao del Rei,
Ouro Branco-MG, 36420-000 Brazil, \\
$^2$ Instituto de F\'isica, Universidade de S\~ao Paulo, 
S\~ao Paulo-SP,  05314-970, Brazil
}
\date{\today}

\begin{abstract}

We investigate the influence of particle diffusion in the two-dimension contact process (CP) with a competitive dynamics
in bipartite sublattices, 
proposed in [Phys. Rev. E {\bf 84}, 011125 (2011)]. 
The particle creation depends on
its first and second neighbors and the extinction increases according to the local density. 
In contrast to the standard CP model, mean-field theory and numerical simulations predict
three stable phases: inactive (absorbing), active symmetric and active asymmetric,
signed by distinct sublattice particle occupations. Our results from MFT and Monte Carlo 
simulations reveal  that low diffusion rates do not destroy sublattice ordering, ensuring
the maintenance of the asymmetric phase. On the other hand, for diffusion
larger than a threshold value $D_c$, the sublattice ordering is suppressed and
only the usual active (symmetric)-inactive transition is presented.  We also show the critical
behavior and universality classes are not affected by the diffusion.
\end{abstract}



\submitto{\JSTAT}

\maketitle

\section{Introduction}

Absorbing-state phase transition manifest themselves when a control parameter
(such as a creation or annihilation rate) is tuned providing
a phase transition from a fluctuating to a state absent of any fluctuation. 
They have deserved considerable interest in recent years, 
being  related to the description of several phenomena  
such as population dynamics, epidemic spreading, chemical reactions and others 
\cite{marro, hinrichsen,odor04,henkel}, as for the search of experimental 
verifications \cite{take07,pine}.

Nowadays it is  widely accepted that absorbing transitions in systems with short-range interactions 
devoid of conserved quantity
or symmetry beyond translational invariance belong to the directed universality (DP) class \cite{gras-jans}.
In another scenario,  the so-called 
DP2 (Z$_2$) universality class embraces systems with  two absorbing states linked by particle-hole symmetry, such
as  branching-annihilating random walks with conserved parity \cite{br}, monomer-monomer reaction models
\cite{mr} and also the voter model \cite{voter}.

Recently, a bidimensional contact process (CP) which exhibits 
sublattice symmetry breaking was proposed 
by de Oliveira and Dickman \cite{martins}.
In addition to the standard creation and annihilation CP mechanisms, 
an activation evolving second-neighbors and
annihilation  depending on the local density are included. 
Besides the usual absorbing (AB) and active [symmetric] (AS) phases, mean field theory (MFT) and Monte Carlo
(MC) analysis predicted the appearance of an unusual active asymmetric 
(AA) phase in which
the distinct sublattices are unequally populated. Remarkably, the symmetric
 and asymmetric phases are separated by a (critical) 
transition presenting  spontaneous symmetry breaking.
 Mean field theory (MFT) and simulations revealed the absorbing phase transition belongs to
the directed percolation (DP) class, whereas the transitions between active phases
fall into the Ising universality class, as expected from symmetry considerations.
The model was extended by Pianegonda and Fiore \cite{pianegonda}, who studied the effects of 
distinct sublattice interactions on the symmetry breaking phase transition.
The changing of interactions can lead to  discontinuous phase transitions
between the absorbing and active phases, although
the criticality is not affected \cite{pianegonda}.

On the other hand, the effects of particle diffusion in competitive contact processes have not been considered yet. 
In particular,
several works have shown that the diffusion can be a relevant perturbation, affecting drastically the
critical behavior \cite{henkel,pcpd,moment} or even leading to distinct scenarios
for discontinuous phase transitions \cite{pianegonda2,fiore-landi,scp}. 
 With these ideas in mind, in the present work, we consider the influence of local diffusion,
aimed at analyzing its effects in another context  than previous studies
(reentrant phase diagram with active phases sharing distinct features,
whose transition is signed by a spontaneous breaking symmetry).

The structure of this paper is organized as follows. In the next
section, we review the model and analyze its mean-field theory.
In Sec. III we present and discuss our simulation results; Sec. IV is
devoted to conclusions.

\section{Model and Mean-Field Theory}

The model  is a stochastic interacting particle system defined on a square lattice, with each site either occupied by a particle
 or vacant. Each particle autocatalytically creates a new particle in one of its first- and second-neighbor empty sites 
with rates $\lambda_1$ and $\lambda_2$, respectively. In a bipartite sublattice, $\lambda_1$ is the rate of creation in the
opposite sublattice, while $\lambda_2$ is the rate in the same sublattice as the replicating particle. Note that unequal sublattice occupancies
are favored for $\lambda_2 > \lambda_1$. An occupied site becomes empty at a rate of unity, independent of the neighboring sites. 
In addition to the intrinsic annihilation rate of unity, there
is a contribution of $\mu n_1^2$, where $n_1$ is the number of occupied
first neighbors.  In order to
understand the meaning of this term in the stabilization
of asymmetric phase, let us consider a scenario
in which the occupation fraction $\rho_A$ of sublattice A
is much larger than that of sublattice B. Particles created in sublattice B
will die out in a short time, stabilizing the unequal sublattice occupancies.
From the above, it is clear that
appropriate quantities for characterizing the phase transitions are densities  of sublattices $A$ and $B$, $\rho_A$ and $\rho_B$, respectively. The total density of particles is given by $\rho=\rho_A+\rho_B$.
In a phase absent of particle creation, we have $\rho_A=\rho_B=0$, consistent with the   absorbing state. 
For distinguishing the  sublattice occupations,  a remarkable quantity is  
$\phi=|\rho_A-\rho_B|$. 
In the absorbing and the active {\em symmetric} (AS) phases, it follows that $\phi=0$, implying that 
there is no difference between the population of sublattices. 
Otherwise, in an active {\em asymmetric} (AA) phase,  $\phi \neq 0$ (since both sublattices are unequally occupied). 
Therefore,  we can use $\phi$ as an order parameter for featuring a {\em spontaneous breaking symmetry}  transition. 

The first inspection over the diffusion effect can be achieved by deriving the one site MFT equations. 
For a lattice of coordination number $q$ ($q=4$ in the square lattice), given by the following coupled equations  

\begin{equation}
\frac{d \rho_A}{dt} = - \left[1 + \mu q^2 \rho_B^2+D\rho_B^*\right] \rho_A + \left[(\lambda_1+D) \rho_B + \lambda_2 \rho_A\right]\rho_A^*
\end{equation}
and
\begin{equation}
\frac{d \rho_B}{dt} = - \left[1 + \mu q^2 \rho_A^2+D\rho_A^*\right] \rho_B + \left[(\lambda_1+D) \rho_A + \lambda_2 \rho_B\right]\rho_B^*
\end{equation}

\noindent where $\rho_A^*=1-\rho_A$ and $\rho_B^*=1-\rho_B$. Note these equations are symmetric under $\rho_A \leftrightharpoons \rho_B$.  
Using the above definitions of $\rho$ and $\phi$, we obtain

\begin{equation}
\frac{d \rho}{dt} = (\Lambda - 1) \rho - \frac{\Lambda}{2} \rho^2 - \frac{\Delta}{2} \phi^2
-\frac{1}{4} \mu q^2 (\rho^2 - \phi^2) \rho,
\label{drhodt}
\end{equation}
and
\begin{equation}
\frac{d \phi}{dt} = \left[ \Delta - 1 - 2D-\lambda_2 \rho -\frac{1}{4} \mu q^2 (\rho^2 - \phi^2)
\right] \phi,
\label{dphidt}
\end{equation}
where $\Lambda \equiv \lambda_1 + \lambda_2$ and $\Delta \equiv \lambda_2 - \lambda_1$.
Eq. (\ref{drhodt}) predicts   the extinction-survival
transition appearing at $\Lambda = 1$, giving rise to the AS phase,  which is linearly stable for small $\Delta$.  
In the AS phase, the stationary  density is given by
\begin{equation}
\rho=\frac{1}{2\kappa}\left[\sqrt{(\Lambda)^2/4+4\kappa(\Lambda-1)}
-(\Lambda)/2 \right],
\label{eqstat1}
\end{equation}
with $\kappa \equiv \mu q^2/4$. From Eq. (\ref{dphidt}), this solution is stable
(while the one with $\phi\neq 0$ is unstable), when
\begin{equation}
a_\phi \equiv \Delta-1-2D-\lambda_2\rho+\kappa \rho^2 < 0,
\end{equation} 
and the transition to the AA phase occurs when $a_\phi=0$.

\begin{figure}[!hbt]
\includegraphics[clip,angle=0,width=1.1\hsize]{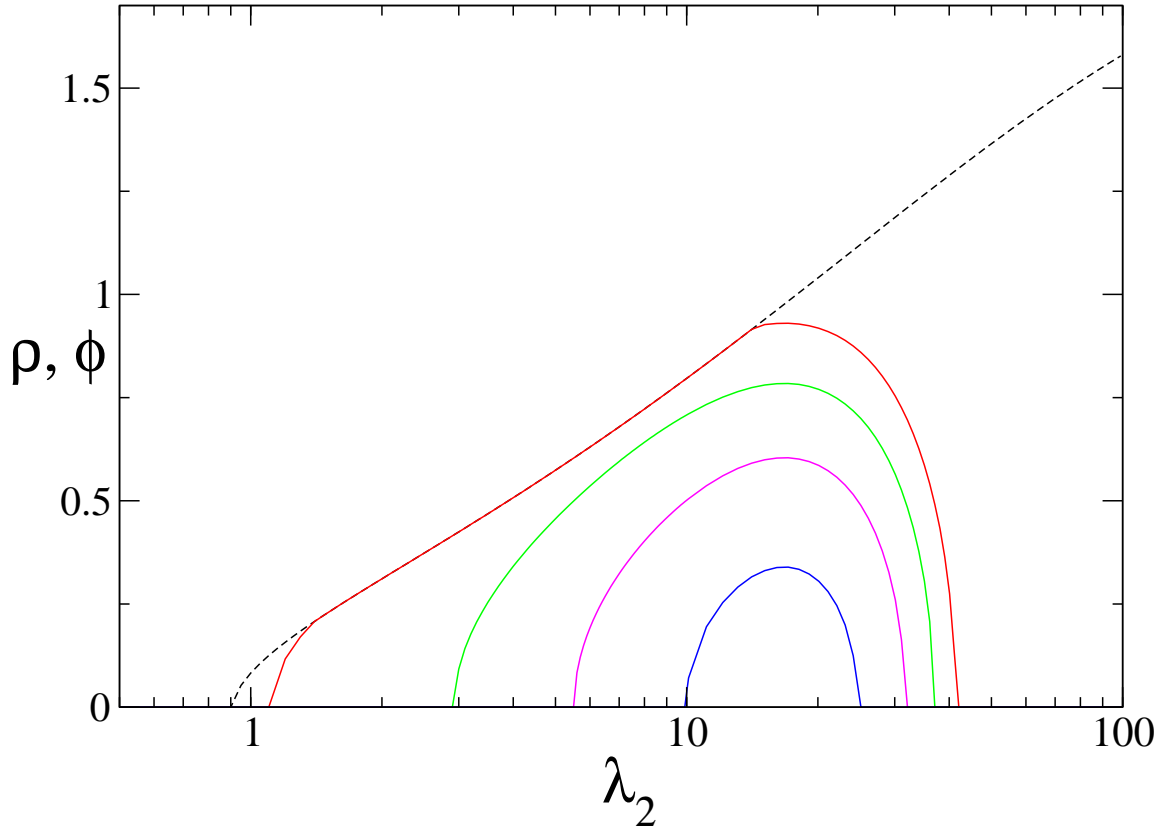}
\caption{\footnotesize{(Color online) Stationary densities of 
$\rho$ (dashed line) and $\phi$ (continuous lines) as function of $\lambda_2$ 
for $\mu = 2$ and $\lambda_1=0.1$. Diffusion rates $D=0, 1, 2$, and 3, from top to bottom.}}
\label{rhophi}
\end{figure}

Eqs. (\ref{drhodt}) and (\ref{dphidt}) lead to the emergence of AA phase
transition for intermediate values of creation parameters and  lower
diffusion rates.  It is stable only
for an intermediate range of $\lambda_2$ and $\lambda_1$. 
By further increasing $\lambda_2$ (for $\lambda_1$ fixed), 
both sublattices become
majority occupied, engendering  a transition from the AA to the AS phases. 
Fig. \ref{rhophi} shows the behavior of both control
parameters $\rho$ and $\phi$ exemplifies the above main features
for $\lambda_1=0.1$ and $\mu=2$. We observe that the AA phase is
strongly dependent on the diffusion rate, in which its range decreases
by raising $D$. Another point concerns that for small $D$, the global
density mildly changes in the AA phase, whereas it exhibits a
monotonous increasing behavior
for the diffusion rates.

In Fig. \ref{phaseD} we show the phase diagram obtained via the MFT 
(continuous lines), for fixed $\lambda_1=0.1$. 
As expected,  low values of  $\lambda_2$ constrains the system
trapped in the absorbing phase, regardless the  diffusion value.  By
increasing $\lambda_2$ the system undergoes a phase transition
from the inactive to the active
symmetric (AS) phase. Similar results are found for other
values of $\lambda_1$. Note that the AA phase decreases by
raising $D$  and disappears
at $D_{MFT}=3.47(1)$, giving rise only to the AS phase. On the contrary,
the absorbing-AS transition line exists for all values of $D$.
In all cases, MFT predicts  continuous phase transitions.

It is important to note that this {\em one-site} MFT neglects all
correlations between nearest-neighbor
sites. In the model, the inhibition term depends  on the local
density, but in the MFT it
appears depending only on the global density. Therefore, the
contribution of the inhibition term is more significant
(which plays an important role in a sublattice
ordering)  for a larger range of parameters than for the lattice
two-dimensional version. 

Although MFT provides a correct {\em qualitative} description of the model,  in the
following section, we perform numerical simulation to compare 
phase diagram and critical properties.

\section{NUMERICAL SIMULATIONS}

\subsection{Methods}

In this work, we performed extensive Monte Carlo simulations of the model on square
lattices of linear size $L=20, 40,..., 320$ sites, with periodic boundary conditions.
The simulation algorithm is the following. First, a
site is selected at random. If the site is occupied, it creates a particle at one of its
first-neighbors with a probability $p_1=\lambda_1/W$, or
at one of its second-neighbors with a probability $p_2=\lambda_2/W$, being $W=(1+\lambda_1+\lambda_2+\mu n_1^2+D)$  the sum of
the rates of all possible events. With a probability
$p_3=D/W$ one of the first neighbor sites is chosen at random and the particle hops to it, provided it is empty. 
Finally, the above-chosen site is vacated with a complementary probability $1-(p_1+p_2+p_3)$, 
in the case it is occupied
\footnote{In order to improve efficiency the sites are chosen from a list which contains the currently
$N_{occ}$ occupied sites; we increment the time by $\Delta t= 1/N_{occ}$ after each event}.
For simulations in the subcritical and in the critical absorbing regimes,
we sample the quasi-stationary (QS) regime using the simulation method detailed in \cite{qssim},
to further improve efficiency.

\subsection{Results and Discussion}
\begin{figure}[!hbt]
\includegraphics[clip,angle=0,width=1.1\hsize]{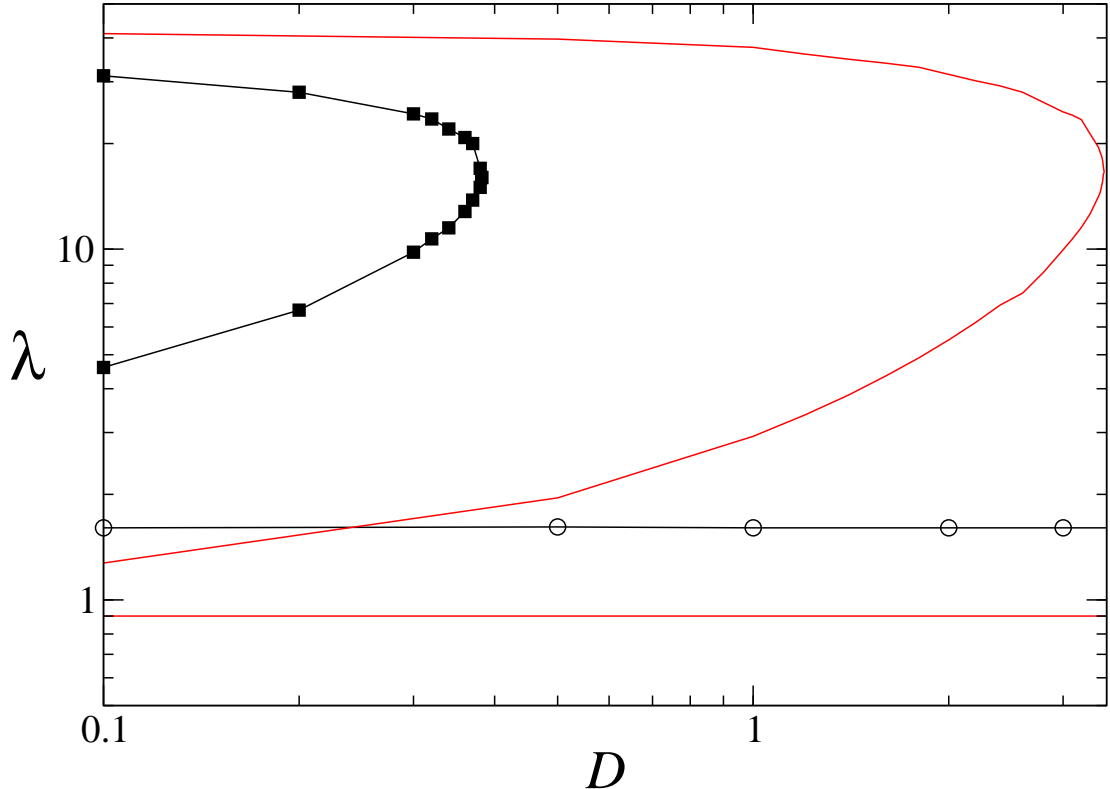}
\caption{\footnotesize{(Color online) Phase diagram in the $D - \lambda_2$ plane
for $\mu = 2$ and $\lambda_1=0.1$, showing absorbing (ABS), active-symmetric (AS) and active asymmetric (AA) phases. Solid lines (red): results from MFT.
Circles: simulation results for the critical points of the absorbing phase transition (black continuous lines are a guide to the eyes).
Solid squares: simulation results for the critical points in the AS-AA boundaries (black continuous lines are a guide to the eyes). In the simulations, the critical points are obtained from an extrapolation for $L\to \infty$ from system sizes up to $L = 320$.}}
\label{phaseD}
\end{figure}

As in the MFT, in all cases, numerical results will be obtained for $\mu=2$.
 Also, the AA phase decreases with the increase of diffusion.
Numerical simulations (see Fig.~\ref{phaseD}) also
show the asymmetric-active phase  only for intermediate 
values of $\lambda_2$, i.e.,
the phase diagram is reentrant. Despite the 
{\em qualitative} agreement between the phase diagram, 
MFT overestimates the regions in 
parameter space corresponding to the
active and ordered phases.  In particular the critical 
diffusion rate (above which the asymmetric-active
phase no longer exists)  
obtained from  MFT $D^{*}_{mft}=3.47$ is about an order of magnitude 
larger than the numerical estimate $D^{*}_{n}=0.382$.

\begin{figure}[!hbt]
\includegraphics[clip,angle=0,width=1.1\hsize]{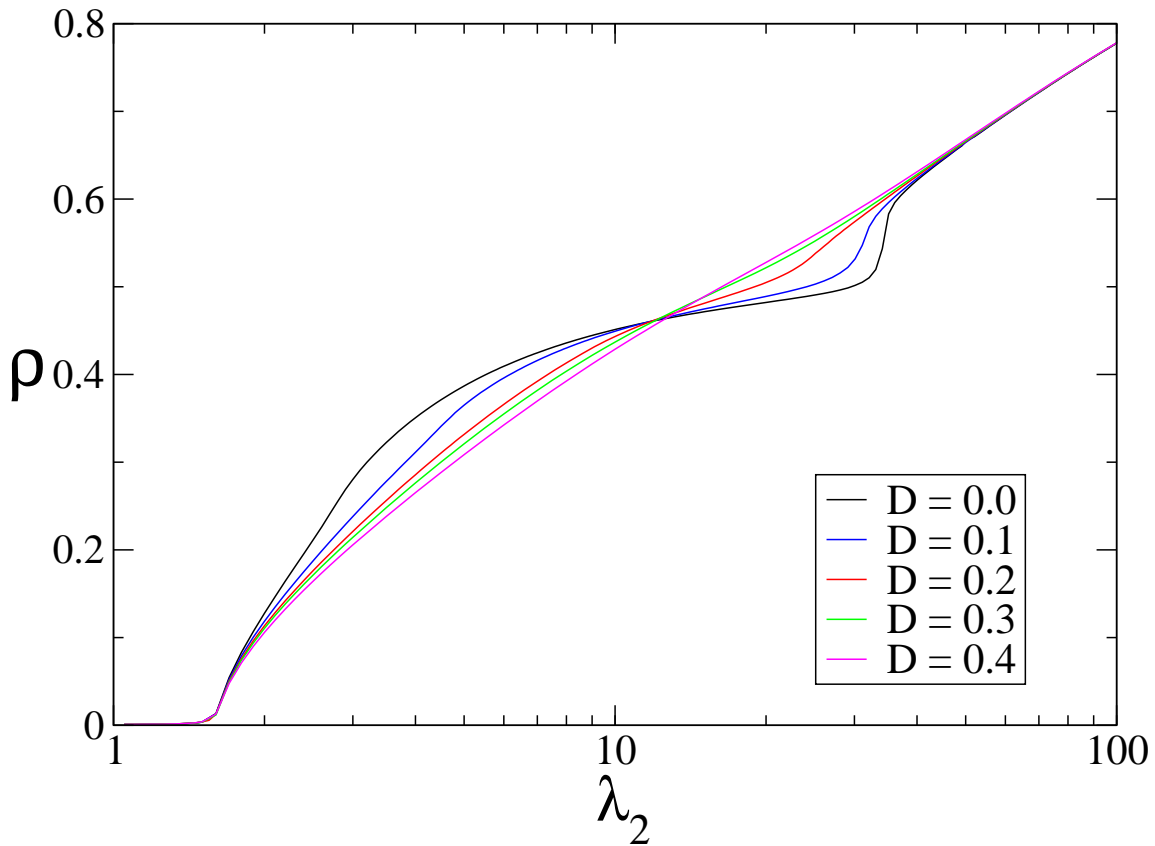}
\caption{\footnotesize{(Color online) Density of active sites $\rho$
for $\mu = 2$ and $\lambda_1=0.1$. Linear system size $L=160$.}}
\label{rhoD}
\end{figure}

\begin{figure}[!hbt]
\includegraphics[clip,angle=0,width=1.1\hsize]{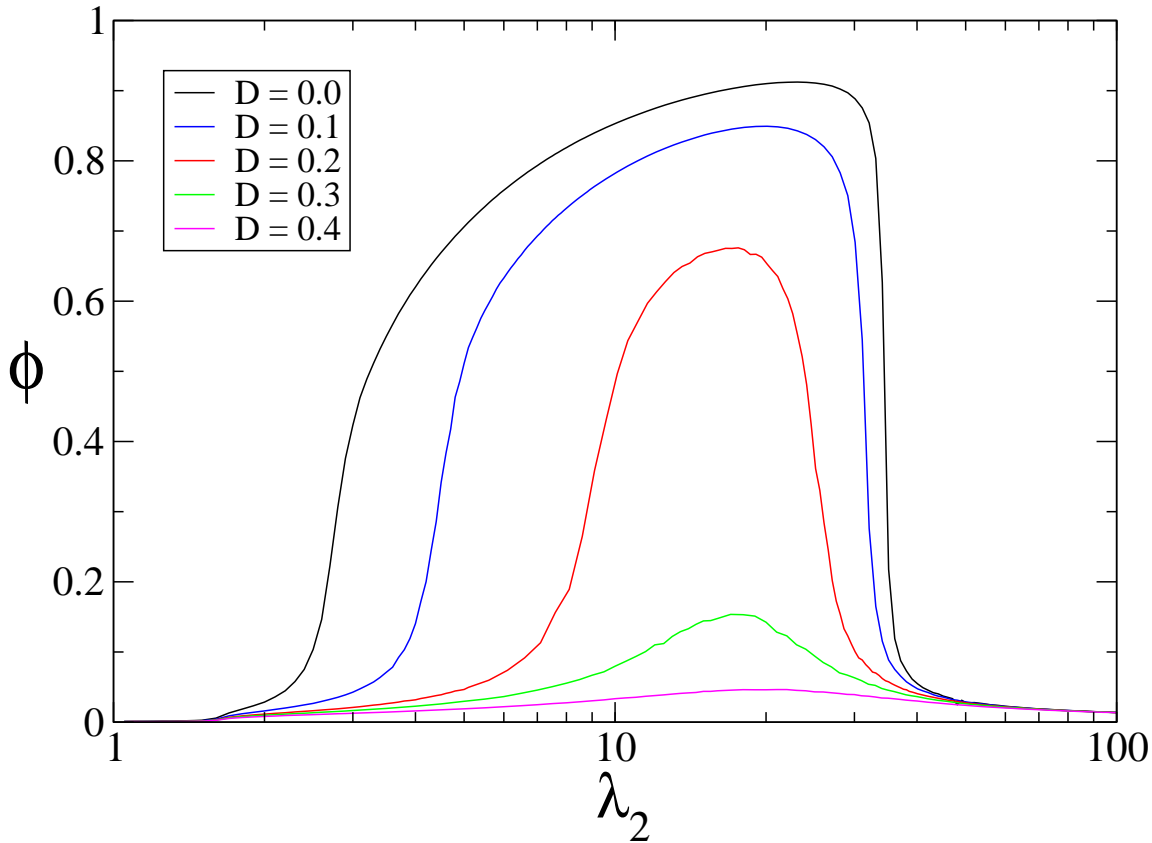}
\caption{\footnotesize{(Color online) Density of parameter $\phi$
for $\mu = 2$ and $\lambda_1=0.1$. Linear system size $L=160$.}}
\label{phiD}
\end{figure}

In Figs. \ref{rhoD} and \ref{phiD} we examine in more details the  main
features of the phase diagram by inspecting the 
order parameters $\rho$ and $\phi$ 
as a function of diffusion. For $D=0$,  in the active phase, $\rho$ 
grows by increasing $\lambda_2$ until its 
saturation very close to $\rho=0.5$ (at $\lambda_2\sim 32.4$).
This behavior is followed by a maximum of $\phi$. Such behavior is a consequence of the inhibition term,  $\mu n_1^2$, that increases with $\phi$ and 
 compete with  the opposite sublattice  term $\lambda_2$. 
By increasing $\lambda_2$ again, 
the creation in the opposite sublattice becomes 
stronger than the inhibition, so that $\rho$ 
 grows faster and $\phi$ decreases towards the vanishing.
 Adding diffusion,  not only  $\phi$ is reduced, but
   also its maximum. In particular,
   $\phi$ vanishes for diffusion $D>0.4$
 This change
 of behavior  induced by the $D$ is closely related
 to  a monotonous increasing of $\rho$ 
(unlike the mildly change as verified for $D=0$). Thus, both 
MFT and numerical simulations predict the suppression
of AA  phase for sufficient large diffusion rates.

\begin{figure}[!hbt]
\includegraphics[clip,angle=0,width=1.1\hsize]{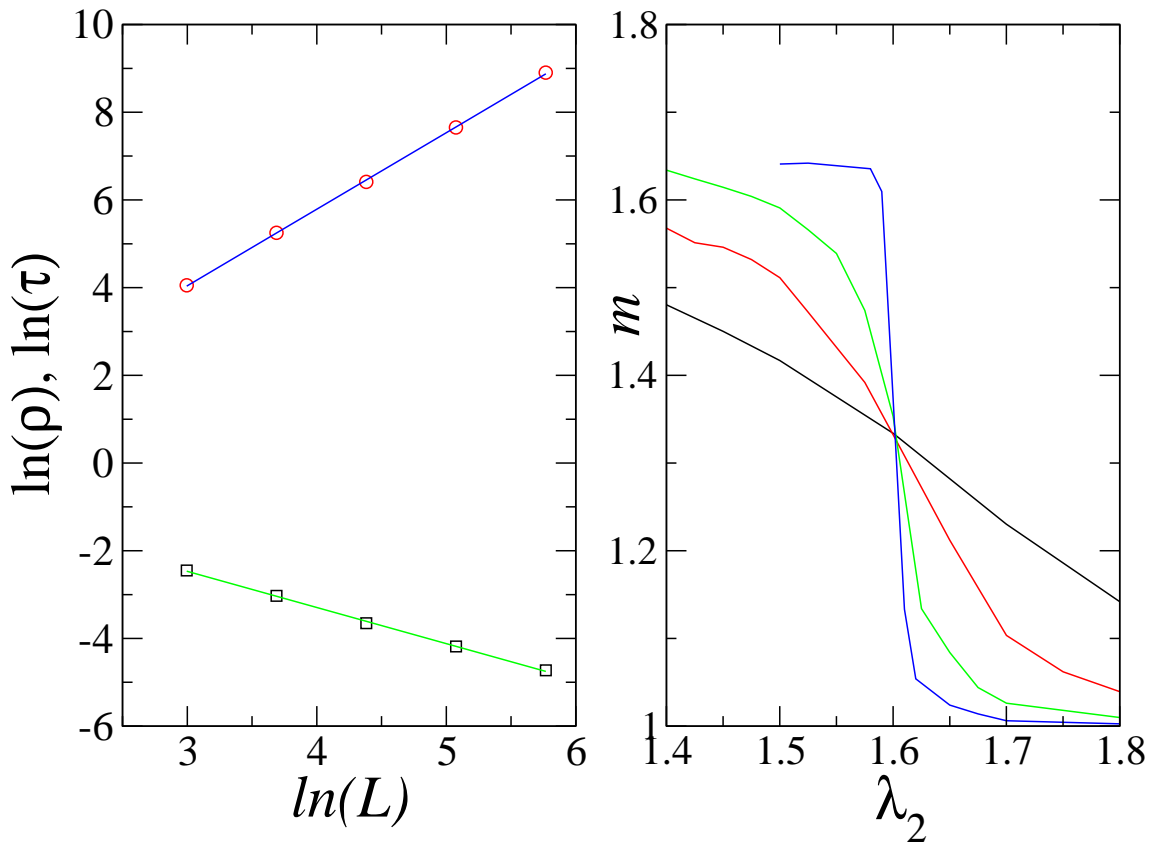}
\caption{\footnotesize{(Color online) (a) Scaled critical QS density of active sites ln $\rho$ (bottom)
and scaled lifetime of the QS state ln $\tau$ (top), {\it
versus} ln$L$. Parameters: $D=0.1$,  $\lambda_1=0.1$ and $\lambda_{2c}=1.6250$ 
(b) Moment ratio $m$ {\it versus} $\lambda_2$. Other parameters same as in (a).}}
\label{exp}
\end{figure}

Now let us examine the critical behavior  for the 
AB-AS, AS-AA and AA-AS phase transitions. 
Starting with the former case, at the critical point the quasistationary 
order parameter  $\rho$ decays as a power law $\rho \sim L^{-\beta/\nu_{\perp}}$,
 being $\beta/\nu_{\perp}$ its associated exponent point.
  In order to locate the transition,  we examine the moment
ratio $m=\langle\rho^2\rangle/\langle\rho\rangle^2$. At
a critical DP like transition, $m$ assumes a universal value
$m_c=1.3264(5)$ \cite{dic-jaf}.
Results from Fig.~\ref{exp}(a)-(b) (for $D=0.1$ and $\lambda_1=0.1$),
reveals that  curves cross at  $\lambda_{2c}=1.6250(5)$ for $m_c=1.330(5)$,
very close to the above DP value. Also,
we obtained $\beta/\nu_\perp=0.81(1)$,  in good agreement with the DP value
$\beta/\nu_\perp=0.797(3)$ \cite{dickman99}. 
For completeness, we also evaluate the
behavior of the lifetime $\tau$  of the QS state at the criticality,
in which an behavior of type
$\tau \sim L^{\nu_{||}/\nu_\perp}$ is expected. From Fig.~\ref{exp}(a), 
we obtain $\nu_{||}/\nu_\perp=1.74(2)$ 
at $\lambda_{2c}$, also in very good agreement with 
the best DP value $\nu_{||}/\nu_\perp=1.7674(6)$.
Therefore we conclude that the absorbing transition belongs to the DP
universality class, as expected.

\begin{figure}[!hbt]
\includegraphics[clip,angle=0,width=1.1\hsize]{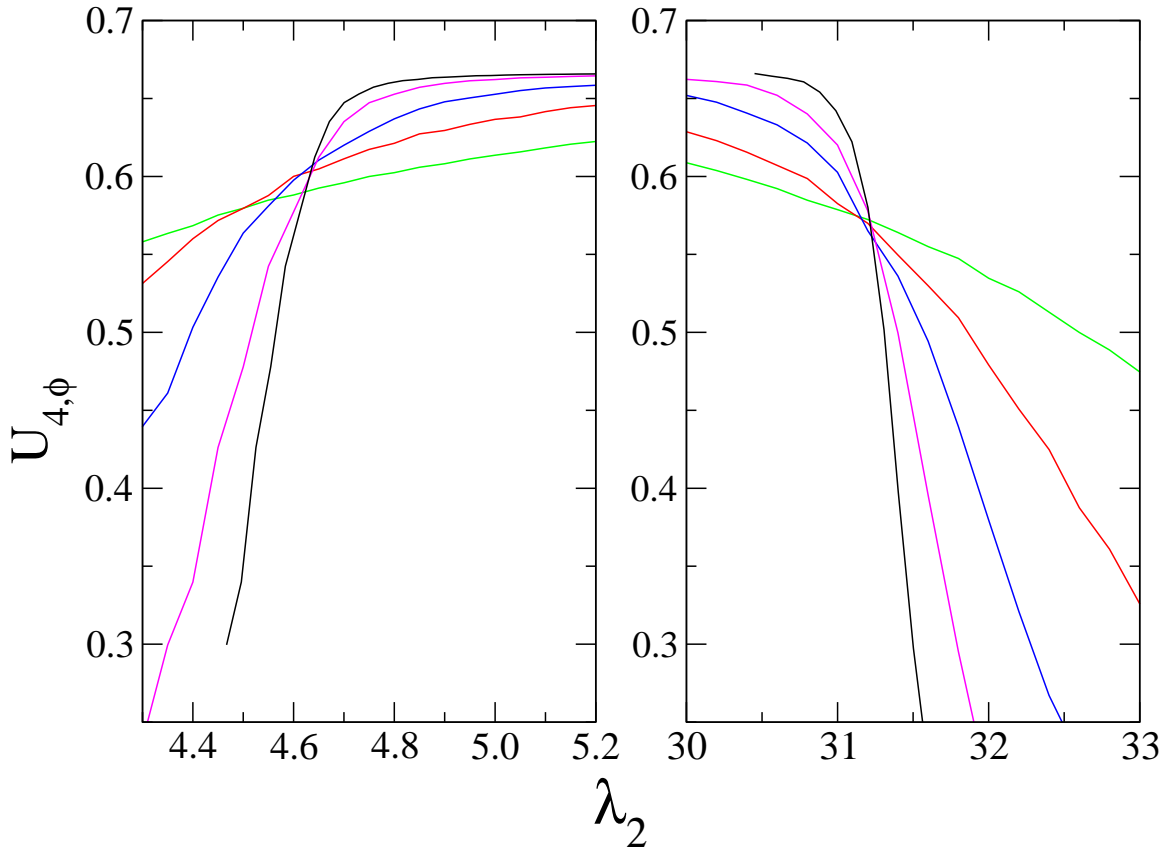}
\caption{\footnotesize{(Color online) Binder cummulant $U_4$ versus $\lambda_2$
for  $\lambda_1=0.1$ and $D=0.1$ for the AS-AA (left) and AA-AS (right)
phase transitions, respectively. System sizes: $L=20,40,80,160$ and $320$.}}
\label{binder}
\end{figure}

Fig. \ref{binder}  exemplify  AS-AA and AA-AS phase transitions
  (also for  fixed $\lambda_1=0.1$) in which a spontaneous-breaking symmetry
is expected.  
For locating the critical point, we
take the reduced Binder cummulant given by \cite{bind81} 
\begin{equation}
U_4=1-\frac{\langle{\phi^4}\rangle}{3\langle{\phi^2}\rangle^2}.
\end{equation}
The intersection points of $U_4$ for successive pairs of sizes
depend rather weakly on the sizes,
providing a reliable estimate for the critical point and approaching to an universal value as $L\to \infty$.
For $\lambda_1=0.1$ and
$D=0.1$, the curves for different sizes intersect at $\lambda_2 = 4.650(5)$, 
and again at $\lambda_2 = 31.27(6)$ when $L\to \infty$,  respectively. 
For the former transition, we found the value
$U_{4,c}=0.615(10)$, very close to the universal value $U_{4,c} =0.61069...$ \cite{kami93}
for the two-dimensional Ising model with fully periodic boundary conditions. 
On the other hand, the latter transition is signed by a significant smaller value of $U_{4,c}=0.55(1)$, 
but close to the value reported by Vasquez and Lopez for the general 
voter model (GVM) \cite{votermodel}. 
For values of $\lambda_2$ between the 
transitions points, the cummulant approaches
2/3, and vanishes outside the asymmetric phase, as 
expected in an ordered phase that 
breaks a  up-down (${\cal Z}_2$) symmetry.

\begin{figure}[!hbt]
\includegraphics[clip,angle=0,width=1.15\hsize]{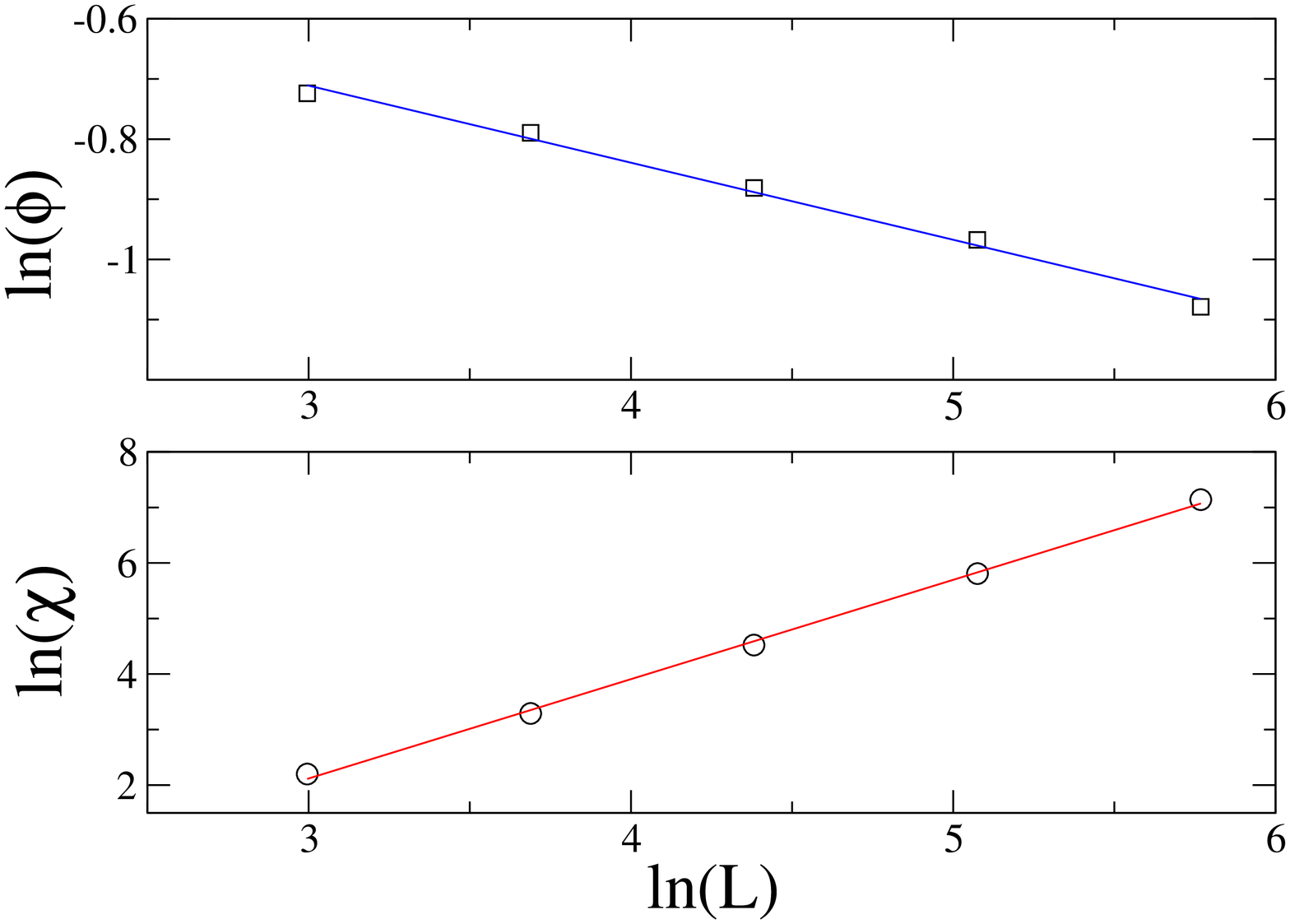}
\caption{\footnotesize{(Color online) Finite size scaling for the critical order parameter $\phi$ (top)
and  maxima of  $\chi$ (bottom). Parameters: $D=0.1$,  $\lambda_1=0.1$ and $\lambda_{2c}=4.65$. }}
\label{fss465}
\end{figure}

 For obtaining the critical exponents, we measure   $\phi$
  and its variance $\chi=L^d(\langle\phi^2\rangle-\langle\phi\rangle^2)$
  for both AS-AA and AA-AS phase transitions.
In all cases, power laws behaviors of type
$\phi \sim L^{-\beta/\nu}$ and $\chi \sim L^{-\gamma/\nu}$ are also expected
at the critical point,
being $\beta/\nu$ and $\gamma/\nu$
the associated critical exponents, respectively.
As shown in Fig.~\ref{fss465}, for the former transition
we obtained $\beta/\nu=0.128(5)$ and  $\gamma/\nu=1.78(4)$,
in good agreement with the exact value
$\beta/\nu=1/8$ and $7/4$ for the Ising universality class
in $d=2$ \cite{bind81}.

\begin{figure}[!hbt]
\includegraphics[clip,angle=0,width=1.15\hsize]{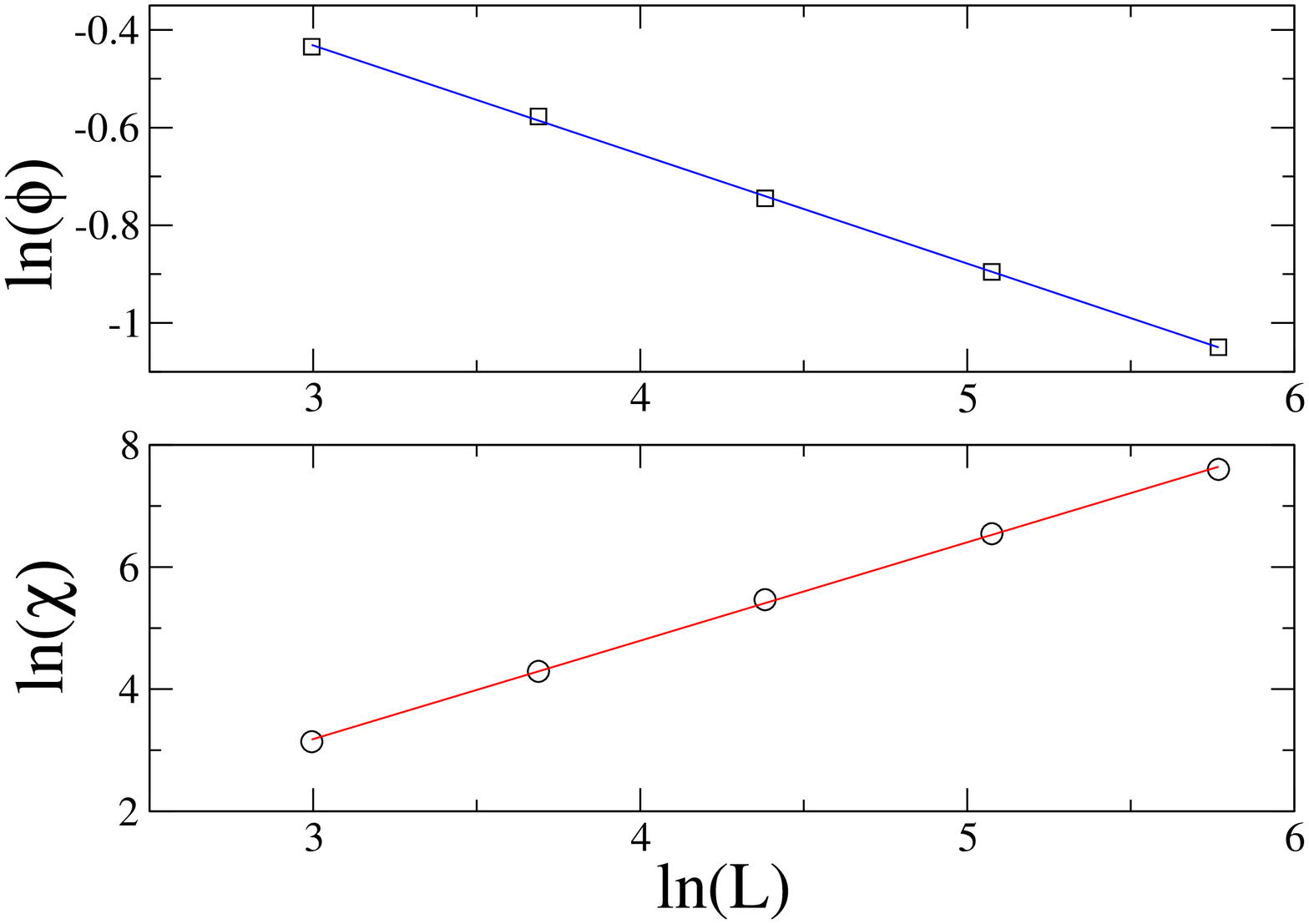}
\caption{\footnotesize{(Color online) Finite size scaling for the critical order parameter $\phi$ (top)
and  maxima of  $\chi$ (bottom). Parameters: $D=0.1$,  $\lambda_1=0.1$ and $\lambda_{2c}=31.27$. }}
\label{fss3127}
\end{figure}

Analysis of AA-AS transition are shown in  Fig.~\ref{fss3127}. 
In this case, we obtain $\beta/\nu=0.21(2)$ and $\gamma/\nu=1.62(4)$
at $\lambda_{2c}=31.27(5)$. These exponents differ from those obtained for
the AS-AA transition, but still obey the scaling relation
$\gamma/\nu=d-2\beta/\nu$.

\section{Conclusions}

In this work, we studied, under mean-field theory and extensive simulations, the influence of 
species diffusion in the phase diagram and critical properties
of  a contact process living in bipartite sublattices.
We observe that low diffusion does not forbid the broken-symmetry
phase with sublattice ordering; however it induces a decreasing
of asymmetric phase. Further increasing the 
diffusion $D$, there is a threshold value, $D^*$,
above which no sublattice order occurs. The phase transitions 
between symmetric-active 
and absorbing phases belong to the 
universality class of directed percolation, whereas the first symmetry-breaking transition is found to be 
Ising-like and the second one seems to belong to the general voter model universality class (this latter transition becomes discontinuous when $D\to 0$) . Thus, despite
leading to remarkable changes in the phase diagram,
the critical behaviors are preserved for the influence of diffusion.
 
Interesting extensions of the present work include the study of the model in disordered environments. 
Disorder can also induce the appearance of spatial \cite{adr-dic,dcp} and temporal Griffiths 
phases \cite{munoz2011}, in which the
disordered system exhibit a different behavior than its pure counterpart \cite{neto,temporal}. 
We believe that an important question to be investigated is 
if such kinds of disorder favor the symmetry-breaking and if it changes the nature of the transitions.

\vspace{1cm}

\vspace{1cm}

\noindent{\bf Acknowledgments}

This work was supported by CNPq, FAPEMIG and FAPESP Brazil. We acknowledge the anonymous referees for their valuable suggestions.

\bibliographystyle{apsrev}

\newpage

\end{document}